\documentclass[11pt]{article}

\usepackage[margin=1in,footskip=0.25in]{geometry}
\usepackage{graphicx}
\usepackage{amsmath}
\usepackage{amssymb}
\usepackage{color}		

\title{Micromorphic effects in an octet truss lattice} 
\author{K. Goyal$\dagger$, R. S. Lakes*\\
* Department of Materials Science, Department of Engineering Physics\\ 
Department of Mechanical Engineering\\
University of Wisconsin\\
1500 Engineering Drive, Madison, WI 53706-1687\\
$\dagger$ Corning Incorporated, Corning, NY, 14830}

\begin{document}
\maketitle

\begin{abstract}
Elastic wave dispersion is studied in an octet truss lattice and compared with a designed rib lattice known to exhibit strong Cosserat elastic effects. Dispersion entails variation of wave speed with frequency.  The phenomenon is experimentally investigated by exciting standing waves in specimens of different length at discrete frequencies.  At lower frequencies corresponding to long wavelengths, wave propagation is classically non-dispersive.  As wavelength approaches a small multiple of the rib length, dispersion is observed. The material exhibited cut-off frequencies above which no signals were propagated.  The physical origin of the dispersion and cut-off is resonance of the ribs. Interpreted as micromorphic continua, cellular solid behavior reveals elastic constants associated with flexibility of the unit cell in comparison with that of the overall material. 
\end{abstract}

\section{Introduction}
Classically elastic materials propagate waves with velocity independent of wavelength or frequency (non-dispersively) and they exhibit bending and torsion rigidity of rods proportional to the fourth power of the diameter \cite{Sokolnikoff}. Materials with microstructure are observed to deviate from classical  elasticity under some conditions. The deviations can be understood via a structural interpretation or via a continuum theory with more freedom than that in classical elasticity. Cosserat \cite{Cosserat09} (micropolar \cite{Eringen68}) elasticity, for example, allows points to rotate as well as translate, and in microstructure elasticity \cite{Mindlin64}, also called micromorphic elasticity, the points in the continuum translate, rotate and deform. Generalized continuum properties have been calculated for some heterogeneous solids via homogenization analysis assuming Cosserat \cite{Askar68} \cite{Tauchert70} or micromorphic \cite{Huttermicro23} response. Nonclassical size effects in which quasistatic torsion and bending rigidity become larger than classical elastic predictions for slender specimens have been observed. These effects have been interpreted via Cosserat elasticity in bone, foams \cite{CossFoam16} and lattices \cite{CossStrong18} \cite{CossGyroid20}. Micromorphic elasticity has sufficient additional freedom to accommodate wave dispersion of shear and longitudinal waves. 
\par
It is well known that in solids with periodic structure, the velocity of the elastic waves depend on the wavelength \cite{Brillouin53}  \cite{Brillouin60}. This phenomenon is called dispersion. At sufficiently high frequency, waves do not propagate: a cut-off frequency is observe. Dispersion of waves of length approaching the scale of atomic spacing in crystals is well known \cite{Nilsson71}. Homogeneous materials, which can be described by classical elastic continuum theory, do not exhibit dispersion.  In such materials, the wave speed remains constant irrespective of the wavelength. Dispersion of waves in periodic particulate composites  \cite{KinraKer83}, materials with a periodic distribution of cavities \cite{AchenbachCav87},  and polymer foams \cite{ChenLakes} is known. 
 Dispersion is usually observed when the wavelength approaches the size of spacing  of structural elements or the frequency of the input signal approaches the resonant frequency of structural elements of the lattice; a cut-off frequency is observed in such materials \cite{Brillouin53}  \cite{Brillouin60}.  The variation in wave velocity can be captured by a mathematical model that incorporates structural resonance. 
\par
\begin{figure}[!htb]
\begin{center}
\includegraphics[width=0.80\textwidth]{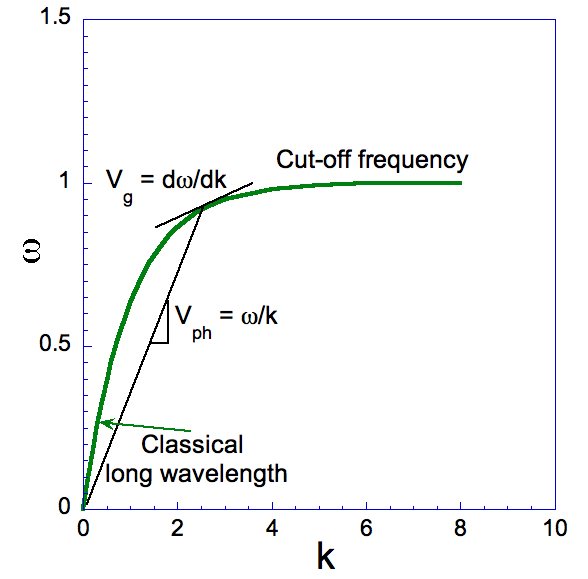}
\caption{A representative graph of angular frequency $\omega = 2 \pi f$ versus wave number $k = \frac{2 \pi}{\lambda}$, with $f$ as frequency and $\lambda$ as wavelength.}
\label{fig:wavedispersion}
\end{center}
\end{figure}

In dispersive media, the phase velocity and group velocity can differ. Phase velocity of waves is the velocity with which the phase front of a single wave moves. 
The phase velocity is given by $V_{ph} = \frac{\omega}{k}$ 
in which  $k = \frac{2 \pi}{\lambda}$ is the wave number with $\lambda$ as wavelength and $\omega = 2 \pi f$ is the angular frequency with $f$ as frequency.
The group velocity is the velocity with which the amplitude modulated wave, or the wave packet, propagates through the material.  The group velocity is given by $V_{g} = \frac{d\omega}{dk}$. 
The group velocity is illustrated by the tangent to the curve while wave velocity is given by the slope of the tangent to the curve in Figure \ref{fig:wavedispersion}. When the group velocity increases with the wave number, it is called positive dispersion. When the group velocity decreases with wave number, it is called negative dispersion \cite{waveprocess}. 
\par
Positive dispersion can arise due to viscoelastic loss or (for shear waves but not longitudinal waves) in Cosserat \cite{Cosserat09} (with inertia terms called micropolar \cite{Eringen68}) elastic effects. Negative dispersion (Figure \ref{fig:wavedispersion}) can arise due to micro-vibration of structural elements in the material, as observed in polymeric cellular materials \cite{ChenLakes} and can be interpreted via microstructure elasticity \cite{Mindlin64}; it does not occur in Cosserat solids.

Wave dispersion has been analyzed in periodic heterogeneous materials \cite{waveElachi76}. Recently, heterogeneous materials designed to control wave propagation have been called acoustic metamaterials \cite{CummerMeta16}, though the concept is anything but new. For example, in membranes with periodic structure \cite{LombMembr} analysis of a periodic spring mass system was done, similarly to analyses done for lattices of atoms. 
\par
In this research, this phenomenon of dispersion is experimentally investigated in a lattice, an octet truss made with Ti-5553 titanium alloy and is compared with a polymer lattice designed for quasi-static nonclassical Cosserat elastic behavior.  A resonance approach is used in which standing waves are excited in the material.      Results are interpreted in the context of the lattice structure and micromorphic elasticity, a generalized continuum.

\section{Materials and Methods}
The material studied was a Ti-5553 titanium alloy octet truss lattice \cite{GoyalRuLak} and is compared with a polyamide polymer lattice designed to exhibit strong nonclassical elastic behavior \cite{CossStrong18}. 
 The octet cell size is approximately 4.5 mm; rib length is 3.2 mm; rib thickness is about 0.53 mm. The polymer lattice contains layers of triangular cells with side length 10.5 mm and layer spacing 9.0 mm. 
\par
The elastic wave dispersion was experimentally investigated by exciting standing waves in the specimen at several values of the fundamental frequency. This was achieved by reducing the specimen length for each test, so that the fundamental frequency of the longitudinal mode of vibration under free-free conditions changes. For the fundamental longitudinal mode in a bar, the wavelength $\lambda$ is twice the specimen length $L$. 
\par
For larger metal octet specimens, the longitudinal mode of vibration was excited via an impulsive tap. The response was obtained by detecting  the emitted sound with a microphone. The specimen was supported by a flexible string at the center, which is a node for the fundamental. This approach suffices to determine the fundamental natural frequency provided it is in the audible range. When the natural frequency was higher, resonant ultrasound spectroscopy (RUS) \cite{LeisureRUS97} \cite{RUS-MiglioriMayn05} was used. This method entails use of piezoelectric transducers to generate and detect ultrasonic signals in a compact specimen. Contact between specimen and transducer was at the corners of the specimen to approximate free vibration conditions by minimizing the effect of transducer rigidity. 
\par
Wave transmission in the lattices was also explored by exciting pulsed waves by an ultrasonic transducer at one end and observing the transmitted signals with a similar transducer at the other end.  Signal quality was insufficient for interpretation.  At the higher frequencies, signal quality was likely impeded by coupling with higher modes. Also the polymer lattice exhibited attenuation of signal due to viscoelastic damping. This was more pronounced at the higher frequencies.  Data points were therefore not shown, but it was possible to ascertain a cut off frequency from the weak signals. 

\section{Results}
\begin{figure}[!htb]
\begin{center}
\includegraphics[width=0.80\textwidth]{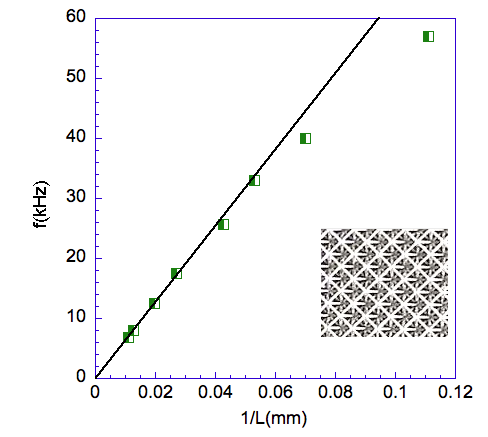}
\caption{Elastic wave dispersion, frequency $f$ vs. inverse length $1/L$, in a Ti-5553 titanium alloy octet lattice. The specimen length is $L = \lambda / 2$. Inset: image of lattice structure; the cell size is 4.5 mm. }
\label{fig:gTiLattice}
\end{center}
\end{figure}
The results for the titanium alloy octet lattice are shown in Figure \ref{fig:gTiLattice}. The black solid line shows the classical asymptote for long wavelength.  The shortest specimen was two cells long. The longitudinal mode of vibration in a specimen one cell long could not be discriminated from a multiplicity of other modes.  In any case, such short specimens transgress the anticipated limits of a generalized continuum theory.  
The results for the octet lattice in Figure \ref{fig:gTiLattice} indicate that the wave velocity decreases as the wavelength decreases. The roll off begins near 30 kHz frequency but the decrease in the slope and hence the group velocity is  gradual. No horizontal asymptote was observed;  the observed cut-off frequency in wave transmission experiments reported for this lattice \cite{GoyalRuLakDamp} began near 60 kHz with full blocking of waves above 100 kHz. 

\begin{figure}[!htb]
\begin{center}
\includegraphics[width=0.80\textwidth]{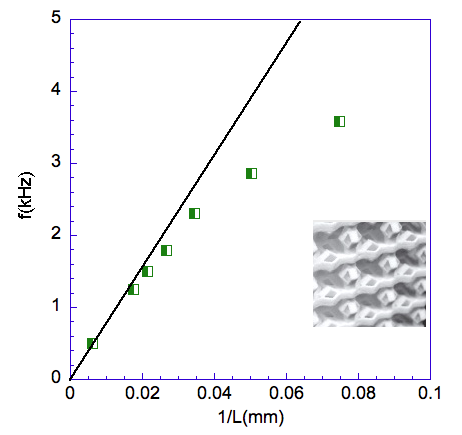}
\caption{Elastic wave dispersion, frequency $f$ vs. inverse length, in a polymer lattice with designed hollow ribs.  The specimen length is $L = \lambda / 2$. Inset: image of lattice structure; the cell size is 9 mm in the longitudinal direction. Adapted from \cite{CossStrong18}. }
\label{fig:PolyLatticDispers}
\end{center}
\end{figure}
Figure \ref{fig:PolyLatticDispers} shows, for comparison, dispersion in a polymer lattice \cite{CossStrong18} designed to exhibit strong nonclassical Cosserat behavior (size effects exceeding a factor of 30) in the quasi-static regime. As above, the black solid line shows the classical asymptote for long wavelength. Size effects refer to deviations of the observed structural rigidity in torsion or bending from the classical dependence of rigidity on the fourth power of the specimen width $w$,  $w^4$.  The smallest specimen was one cell in length. In contrast to the octet lattice the polymer lattice displays, in Figure \ref{fig:PolyLatticDispers}, a more pronounced roll off of the velocity. Even so, the curve did not approach the horizontal branch illustrated in Figure \ref{fig:wavedispersion}. The horizontal branch was, by contrast, observed in polymer foams \cite{ChenLakes} with positive and negative Poisson's ratio. For these foam materials, the shortest specimens were many cells long  so a continuum interpretation is more reasonable.

\section{Analysis}
\subsection{Microstructure / micromorphic elasticity}
In microstructure elasticity  \cite{Mindlin64}, also called micromorphic elasticity, the points in the continuum translate as they do  in classical elasticity, rotate as they do in Cosserat elasticity, and deform as well. A micro-deformation $\psi_{ij}$ is defined such that the symmetric part $\psi_{(ij)} = \frac{1}{2}(\psi_{ij} + \psi_{ji})$ is the micro-strain and the antisymmetric part $\psi_{[ij]} = \frac{1}{2}(\psi_{ij} - \psi_{ji})$ is the micro-rotation.  The micro-rotation differs from the macro-rotation $r_{i} = (e_{ijk}u_{k,j})/2$ which depends on gradients of displacement as in the Cosserat theory, and $e_{ijk}$ is the permutation symbol. The small strain tensor is ${\epsilon}_{ij} =  \frac{1}{2} (u_{i,j} + u_{j,i})$ in which $u_{i}$ the displacement vector. The strain is called the macro-strain to distinguish it from the micro strain.  The relative deformation ${\gamma}_{ij}$  is ${\gamma}_{ij} = \frac{\partial u_{j}}{\partial x_{j}} - \psi_{ij} $. The macro gradient of micro-deformation, called the micro-deformation gradient, is called $\varkappa_{ijk} = \partial_{i}\psi_{jk}$. 
\par
\begin{figure}[!htb]
\begin{center}
\includegraphics[width=0.7\textwidth]{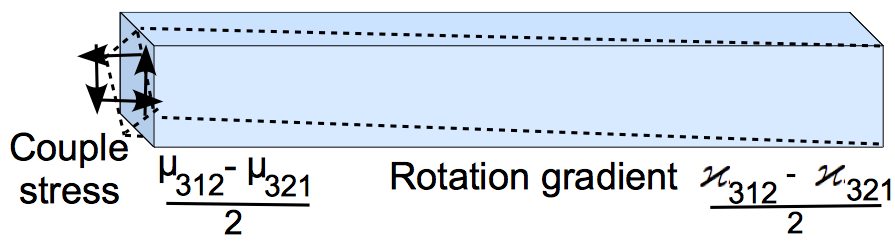}\\
\includegraphics[width=0.7\textwidth]{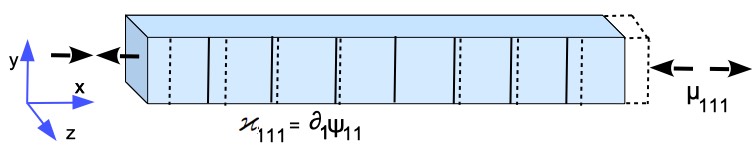}
\caption{Gradients of micro-deformation $\varkappa_{pqr}$ and corresponding double stress $\mu_{pqr}$, adapted from \cite{Mindlin64}. Arrows represent forces. Top: Cosserat freedom within micromorphic framework. Bottom: micromorphic freedom for axial compression. }
\label{fig:Mindlinfig}
\end{center}
\end{figure}

The stress corresponding to the strain ${\epsilon}_{ij}$ is,from the strain energy $W$, the Cauchy stress ${\tau}_{ij} = \frac{\partial W}{ \partial {\epsilon}_{ij}}$. The stress corresponding to the relative deformation ${\gamma}_{ij}$  is called the relative stress ${\sigma}_{ij} = \frac{\partial W}{\partial {\gamma}_{ij}}$. The stress corresponding to  the micro-deformation gradient $\varkappa_{pqr}$ is called the double stress $\mu_{ijk}  = \frac{\partial W}{\partial {\varkappa_{ijk}}}$.  The first subscript of $\mu_{ijk}$ refers to the normal to the surface across which the component acts.  Double stress refers to pairs of forces per unit area. If the pair gives rise to a moment it includes the couple stress of Cosserat elasticity as shown in Fig. \ref{fig:Mindlinfig}. Mindlin  \cite{Mindlin64} provides additional illustrations of the geometry of micro-deformation gradient and double stress.   Plane waves which depend on modulus and Poisson's ratio were not considered because the Poisson's ratio of the lattices was not measured.     	

The constitutive equations for the isotropic microstructure theory are for the Cauchy stress, Eq. (\ref{eq:con1}), for the relative stress Eq. (\ref{eq:con2}) and for the double stress, Eq. (\ref{eq:con3}), in which $\delta_{pq}$ is the Kronecker delta. 
\begin{equation} \label{eq:con1}
\tau_{pq} = \lambda \delta_{pq} + 2G \epsilon_{pq} + g_{1} \delta_{pq} \gamma_{ii} + g_{2}(\gamma_{pq} - \gamma_{qp} )
\end{equation}  
\begin{equation} \label{eq:con2}
\sigma_{pq} = g_{1} \delta_{pq} \epsilon_{ii} + 2g_{2} \epsilon_{pq} + b_{1}\delta_{pq}\gamma_{ii} + b_{2}\gamma_{pq} + b_{3}\gamma_{qp}
\end{equation}  
\begin{align}  \label{eq:con3}
\begin{split}
\mu_{pqr} = a_{1} (\varkappa_{iip}\delta_{qr}  + \varkappa_{rii}\delta_{pq} ) + a_{2} (\varkappa_{iiq}\delta_{pr}  + \varkappa_{iri}\delta_{pq}) + a_{3} \varkappa_{iir}\delta_{pq} + a_{4}\varkappa_{pii}\delta_{qr}\\
+ a_{5} (\varkappa_{qii}\delta_{pr}  + \varkappa_{ipi}\delta_{qr}) + a_{8} \varkappa_{iqi}\delta_{pr} + a_{10} \varkappa_{pqr} + a_{11} (\varkappa_{rpq} + \varkappa_{qrp})+\\
 a_{13} \varkappa_{prq} +  a_{14} \varkappa_{qpr} +  a_{15} \varkappa_{rqp}
\end{split}
\end{align}
Isotropic microstructure / micromorphic elasticity has 18 independent elastic constants (${\lambda}$,  ${G}$, $g_{i}$, $b_{i}$, $a_{i}$) compared with 6 for the isotropic Cosserat solid and 2 for the isotropic classical solid. 

\subsection{Interrelation among symbols}
Different notations used for Cosserat elasticity were compared \cite{CowinCoss70}. The Cosserat coupling constant \cite{Eringen68}   is  $\kappa$ = $2\beta$ in the symbols of Mindlin  \cite{Mindlin65}. This constant governs the coupling between micro and macro rotations and is a measure of the magnitude of effects predicted. 
Correspondences between symbols used for microstructure elasticity \cite{Mindlin64} and those used by Mindlin for Cosserat elasticity were provided in ref. \cite{Mindlin65}. Mindlin's Cosserat $\beta$ = $\frac{1}{2}(b_{2} - b_{3})$ in microstructure elasticity. So the Cosserat coupling constant is $\kappa$ = $(b_{2} - b_{3})$ in microstructure elasticity. 
\par
The Cosserat constants associated with rotation gradient sensitivity \cite{Mindlin65} are linear combinations of several of the microstructure elasticity coefficients including $a_{10}$ and $a_{13}$.

\subsection{Wave effects}
The cut-off frequency was predicted by determining the fundamental natural frequency of a rib element. 
The fundamental frequency $f$ for a simply supported bar \cite{BenVibr} of Young's modulus $E$, density $\rho$, length $L$ and area moment of inertia $I$ in bending is, in which $q$ is the mass per unit length, and $I = \frac{1}{12} w^{4}$, so  $f = \frac{\pi}{2} [\frac{EI}{q L^{4}}]$. For a bar of square cross section thickness $w$, $q = \rho w^{2}L /L$, so
\begin{equation} \label{eq:bendvib1}
f = \frac{\pi}{4\sqrt{3}}  \frac{w}{L^{2}} \sqrt{\frac{E}{\rho}} .
\end{equation}

The ribs are connected to other ribs so the end boundary condition is not as simple as the assumption. The numerical multiplier may therefore differ by a factor of order unity. 
For titanium alloy octet lattices, the rib resonance occurred near 100 kHz. The observed cut-off frequency was near 100 kHz; at 60 kHz there was a substantial reduction in the transmitted signal \cite{GoyalRuLakDamp}. 
\par
The polymer lattice exhibits transmissibility that begins to roll off above about 2 kHz. Analysis of rib resonance would be complicated by the fact the ribs are hollow and have nonuniform cross section.  The experimental results for the polymer lattice indicate a cut-off frequency near 4 kHz. Viscoelastic dissipation in the polymer lattice will give rise to curve that is concave up, but the experiments show that the group velocity is decreasing, which indicates that micro-vibration of the rib elements have a dominant effect in dispersion of the elastic waves in the polymer lattice. 
\par
Lattices such as the octet contain a triangulated structure in which the ribs must deform axially. Such lattices are called stretch dominated \cite{GibsonAshby}. The modulus is given by $\frac{E}{E_{s}} \propto [\frac{\rho}{{\rho}_{s}}]$ in which $E_{s}$ is Young's modulus of the solid ribs,  $E$  is Young's modulus of the lattice,  ${\rho}_{s}$ is the density of the solid ribs and  ${\rho}$ is the density of the lattice. For the octet in principal directions \cite{DeshpandeAshby01}, the constant of proportionality is about 1/9.  The stiffness of the octet has made it attractive in design of turbine blades with a core made of octet lattice; these blades provide improved ratios of strength to weight in comparison with solid blades \cite{OctetTurbine}. The lattice filling also raised the fundamental frequency of vibration.  By contrast, in foams, the ribs bend rather than stretch or compress. The modulus \cite{GibsonAshby} of bend-dominated foam is given by $\frac{E}{E_{s}} \propto [\frac{\rho}{{\rho}_{s}}]^{2}$; for foams of normal structure the constant of proportionality is close to 1. The difference in exponent means foams of low density are more compliant than lattices of the same density made of the same material. 
\par
If it is desirable to block waves at lower frequencies, one can make the cells larger. Increasing the slenderness $w/L$ of the ribs will lower the rib natural frequency as well but will also reduce the stiffness and strength of the lattice; similarly with the rib modulus. If one intends to lower the cut-off frequency and retain the lattice stiffness and strength, only the cell size should be changed. 
\par
Wave dispersion and cut-off frequencies do not occur in classical elasticity. This continuum representation of the material has insufficient freedom. Cosserat elasticity incorporates more freedom, and allows the velocity of shear waves to increase with frequency. There is no predicted dispersion of longitudinal waves in Cosserat solids.  Cosserat elasticity allows points to rotate as well as translate; it admits moments per area (couple stress) and it incorporates 6 elastic constants for an isotropic solid.  Cosserat solids incorporate sensitivity to rotation gradient; they exhibit quasi-static effects such as size effects in torsion and bending as demonstrated experimentally in foams \cite{CossFoam16}, rib lattices \cite{CossStrong18} and surface lattices \cite{CossGyroid20}.  Cosserat solids also exhibit reduction of concentrations of stress and strain  \cite{Mindlin63}; also dynamic effects including dispersion of shear waves in which velocity increases with frequency via continuum theory \cite{Eringen68} and experiment  \cite{Merkel11}.  
\par
Micro-structure / micromorphic elasticity incorporates sufficient freedom in a continuum approach to allow dispersion of longitudinal waves and cut-off frequencies   \cite{Mindlin64}.  The theory incorporates 18 elastic constants for an isotropic solid; it incorporates sensitivity to rotation gradient and strain gradient and includes Cosserat elasticity as a special case. The micro-structure elasticity theory \cite{Mindlin64} gives the angular frequency $\omega$ for micro-vibration for waves in equi-voluminal extensional modes as 

\begin{equation} \label{eq:MindlinMicrVib}
\omega^{2} = \frac{3(b_{2} +b_{3})}{\rho d^{2}}
\end{equation}  

with $\rho$ as density,  $b_{2}$ and $b_{3}$ as nonclassical micromorphic elastic constants Eq. (\ref{eq:con3}) and $d$ as a microstructure size. The elastic constants $b_{2}$ and $b_{3}$  link the relative stress with the relative deformation (the difference between the macro-displacement gradient and the micro-deformation). 
The micro-deformation contains an asymmetric part corresponding to the Cosserat micro-rotation and a symmetric part called the micro-strain which can differ from the macroscopic strain. 
\par
In the context of interpreting experiments \cite{ChenLakes},  a dimensionless measure has been defined, 
\begin{equation} \label{eq:Lambda}
\Lambda^{2} = \frac{\omega_{c}^{2} \rho d^{2} }{3G}
\end{equation} 
with $G$ as the shear modulus and $\omega_{c}$ as the observed cut-off angular frequency. The range for $\Lambda$ is from zero to infinity.  
So the cut-off frequency may be used to infer $b_{2} +b_{3}$ in the micromorphic theory. 
\par
The octet lattice density was 0.534 g/cm$^{3}$ so for a cell size 4.5 mm and a shear modulus 1 GPa,  a cut-off frequency $f$ = 100 kHz results in $\Lambda$ = 1.2 for the octet lattice.
For the strongly nonclassical polymer lattice, the  density was 0.16 g/cm$^{3}$, the shear modulus was $G$ = 1.1 MPa and $d$ = 9 mm. For a cut-off frequency $f$ = 4 kHz, $\Lambda$ = 1.6 for the polymer lattice. 
Consequently the normalized micromorphic effect $(b_{2} +b_{3})/G$ is similar for the two lattices. One may express the dimensionless measure as 

\begin{equation} \label{eq:Lambda2}
\Lambda^{2} = \frac{3(b_{2} +b_{3})}{G} .
\end{equation} 

By contrast, the Cosserat effects differed substantially: the strongly nonclassical polymer lattice \cite{CossStrong18} exhibited size effects of a factor of 36 in torsion and 29 in bending but the octet lattice exhibited a size effect of a only a factor 1.3 in torsion and 1.1 in bending \cite{GoyalRuLak}. 
\par
For foams \cite{ChenLakes}, $\Lambda$ was 0.089 for as-received polyester foam of conventional structure and cell size 0.5 mm, and 0.03 for re-entrant polyester foam with a negative Poisson's ratio. 
Because $\Lambda$ was identified with a nonclassical rigidity of the unit cell in comparison with the rigidity of the bulk material, the foams may be regarded as having compliant unit cells compared with the lattices. The overall modulus of the material is incorporated in the calculation of $\Lambda$, so the compliance of the foam is not responsible for the difference. A detailed analysis of micromorphic elasticity of cellular solids is presented elsewhere \cite{LakesMicromorphic}. 
 
\subsection{Size effects}  
Size effects do not occur in the compression of classical or Cosserat solids, but they can occur in micromorphic solids  \cite{LakesMicromorphic}. Size effects refer to deviations of the observed structural rigidity in torsion or bending from the classical dependence of rigidity on the fourth power of the specimen width $w$,  $w^4$.    Some materials studied in the context of Cosserat freedom were also explored for size effects in compression \cite{Cosslattices19}. For cubic unit cell lattices with $<111>$ cell axes and with  $<100>$ cell axes parallel to the longitudinal axis the observed Young's modulus and Poisson's ratio were independent of width. By contrast, these lattices exhibited substantial Cosserat type size effects in torsion and in bending.     Large size effects were observed in designed lattices \cite{Cosslattices19}.   
\par
Size effects in compression were not observed in the titanium octet truss lattice. 

\section{Conclusions}
Dispersion and cut-off frequencies occur in cellular solids such as rib lattices. The physical structural interpretation is resonance in the rib elements of the materials. The cut-off frequency depends on the cell size, slenderness and rib material. The continuum interpretation of the phenomena is done by micromorphic generalized continuum theory which is more general than Cosserat elasticity.   Both structural and continuum theories account for dispersion and cut off frequencies.   	

\section{Acknowledgment}
 Discussions and support of prior study of lattices by Z. Rueger are gratefully acknowledged.

\end{document}